\documentclass[nobalancelastpage,twocolumn,floatfix,superscriptaddress,aps,prb]{revtex4-2}

\usepackage{microtype}
\usepackage[hidelinks,colorlinks,citecolor=blue,linkcolor=black,urlcolor=blue]{hyperref}
\usepackage{graphicx}
\usepackage{amsmath,amssymb,amsfonts}
\usepackage[capitalize]{cleveref}
\usepackage{xspace}
\usepackage{dsfont}
\usepackage[caption=false]{subfig}
\usepackage{floatrow}

\graphicspath{{./}{../plots/}}

\newcommand{\diff}{\mathrm{d}\xspace}
\newcommand{\tr}[1]{\mathrm{Tr}\left[#1\right]}
\newcommand{\sdiag}{S_{\mathrm{diag}}}

\newsavebox{\measurebox}

\begin{document}

\title{Measurement-induced entanglement transitions in many-body localized systems}

\author{Oliver Lunt}
\email{oliver.lunt.17@ucl.ac.uk}
\affiliation{Department of Physics, University College London, Gower Street, London, WC1E 6BT}

\author{Arijeet Pal}
\email{a.pal@ucl.ac.uk}
\affiliation{Department of Physics, University College London, Gower Street, London, WC1E 6BT}

\date{\today}

\begin{abstract}
The resilience of quantum entanglement to a classicality-inducing environment is tied to fundamental aspects of quantum many-body systems. The dynamics of entanglement has recently been studied in the context of measurement-induced entanglement transitions, where the steady-state entanglement collapses from a volume-law to an area-law at a critical measurement probability~$p_{c}$. Interestingly, there is a distinction in the value of $p_{c}$ depending on how well the underlying unitary dynamics scramble quantum information. For strongly chaotic systems, $p_{c} > 0$, whereas for weakly chaotic systems, such as integrable models, $p_{c} = 0$. In this work, we investigate these measurement-induced entanglement transitions in a system where the underlying unitary dynamics are many-body localized (MBL). We demonstrate that the emergent integrability in an MBL system implies a qualitative difference in the nature of the measurement-induced transition depending on the measurement basis, with $p_{c} > 0$ when the measurement basis is scrambled and $p_{c} = 0$ when it is not. This feature is not found in Haar-random circuit models, where all local operators are scrambled in time. When the transition occurs at $p_{c} > 0$, we use finite-size scaling to obtain the critical exponent $\nu = 1.3(2)$, close to the value for 2+0D percolation. We also find a dynamical critical exponent of $z = 0.98(4)$ and logarithmic scaling of the R\'{e}nyi entropies at criticality, suggesting an underlying conformal symmetry at the critical point. This work further demonstrates how the nature of the measurement-induced entanglement transition depends on the scrambling nature of the underlying unitary dynamics. This leads to further questions on the control and simulation of entangled quantum states by measurements in open quantum systems.
\end{abstract}

\maketitle
    
\section{Introduction}
\label{sec:introduction}

The environment plays a profound role in the dynamics of quantum systems. The loss of coherence of the quantum system is facilitated by the environment which ultimately leads to the emergence of classical behaviour at the macroscopic scale \cite{RevModPhys.76.1267_Schlosshauer}. While the dynamics of an isolated quantum system is unitary, the coupling to the environment renders it non-unitary. Dynamics of quantum entanglement due to non-unitary dynamics is therefore crucial for the understanding of the quantum to classical crossover. A particular framework in which to explore non-unitary dynamics is in the context of quantum measurements. The influence of measurements has shed new light on the dynamics of information in quantum many-body systems and provided potential mechanisms for purification \cite{PhysRevLett.76.722_Bennett, PhysRevA.71.022316_Bravyi} and quantum error-correction \cite{PhysRevA.54.3824_Bennett}. Exciting experimental developments over the last decade in quantum control and simulation of few and many-body systems have shed light on the interplay between measurement, entanglement, and decoherence \cite{PhysRevLett.106.110502_Vijay,murchObservingSingleQuantum2013,ofekExtendingLifetimeQuantum2016,minev2019QuantumJump}. 

Measurement-induced dynamics has led to the unravelling of a new class of entanglement phase transitions driven by quantum measurements \cite{liQuantumZenoEffect2018,skinnerMeasurementInducedPhaseTransitions2019,baoTheoryPhaseTransition2020,chanUnitaryprojectiveEntanglementDynamics2019,jianMeasurementinducedCriticalityRandom2020,szyniszewskiEntanglementTransitionVariablestrength2019,gullansScalableProbesMeasurementinduced2019,liMeasurementdrivenEntanglementTransition2019,caoEntanglementFermionChain2019,zabaloCriticalPropertiesMeasurementinduced2020,tangMeasurementinducedPhaseTransition2020,rossiniMeasurementinducedDynamicsManybody2020,gotoMeasurementInducedTransitionsEntanglement2020,liConformalInvarianceQuantum2020,lopez-piqueresMeanfieldTheoryEntanglement2020,fanSelfOrganizedErrorCorrection2020,zhangNonuniversalEntanglementLevel2020,szyniszewskiUniversalityEntanglementTransitions2020,fujiMeasurementinducedQuantumCriticality2020,sangMeasurementProtectedQuantum2020,lavasaniMeasurementinducedTopologicalEntanglement2020,shtankoClassicalModelsEntanglement2020,vijayMeasurementDrivenPhaseTransition2020} in random unitary circuits, which have served as effective models for quantum chaos and thermalization. The study of these transitions has also led to new insights into emergent quantum error-correction in many-body quantum systems \cite{choiQuantumErrorCorrection2019,gullansDynamicalPurificationPhase2019,fanSelfOrganizedErrorCorrection2020}, and to efficient classical algorithms for the simulation of random shallow 2+1D quantum circuits \cite{nappEfficientClassicalSimulation2019}. The measurements drive the system into a steady state which either satisfies a volume law for entanglement or an area law, depending on how frequently the quantum degrees of freedom are being measured. The transition only occurs in an individual quantum trajectory conditioned on the measurements: averaging over the measurements washes out the distinct entanglement structure in the steady states in the two phases. Although in these toy models of the transition the measurements are projective, analogous transitions are also realized in continuously measured systems.  These 1+1D quantum circuits can be mapped to a 2+0D classical statistical mechanical model, which in the limit of large local Hilbert space dimension maps to a percolation model \cite{skinnerMeasurementInducedPhaseTransitions2019,baoTheoryPhaseTransition2020,jianMeasurementinducedCriticalityRandom2020, liConformalInvarianceQuantum2020}, explaining the emergence of criticality at the transition, and critical exponents which are mostly those of 2+0D percolation.

Although not easily tractable, the study of the effect of measurements on Hamiltonian dynamics has potentially direct relevance to experiments. It is experimentally viable to realise non-integrable quantum Hamiltonians whose effective dynamics is described by random unitary circuits \cite{mandel2003multipartite, kaufman2016quantumthermalization, bernien2017QuantumSimulator, zhang2017manybody}. Systems whose dynamics are governed by an interacting Hamiltonian exhibit two classes of dynamics. At one end are chaotic systems, which can variously be characterized as those systems which spread entanglement ballistically and saturate to volume-law steady state entanglement, complying with equilibrium thermodynamics \cite{dalessioQuantumChaosEigenstate2016}. At the other end are many-body localized (MBL) systems, which violate equilibrium thermodynamics \cite{basko2006MBL, palManybodyLocalizationPhase2010, PhysRevB.75.155111_Oganesyan, nandkishoreManyBodyLocalizationThermalization2015, abaninManybodyLocalizationThermalization2018,abaninRecentProgressManybody,parameswaranManybodyLocalizationSymmetry2018}, and can similarly be characterized by their logarithmic spreading of entanglement in time and also saturate to volume-law steady state entanglement, albeit lower than the thermodynamic entropy density \cite{PhysRevB.77.064426_Znidaric, bardarsonUnboundedGrowthEntanglement2012, husePhenomenologyFullyManybodylocalized2014, serbynUniversalSlowGrowth2013, nanduriEntanglementSpreadingManybody2014}.

\begin{figure}[t]
    \centering
    \sidesubfloat[]{\includegraphics[width=0.9\columnwidth]{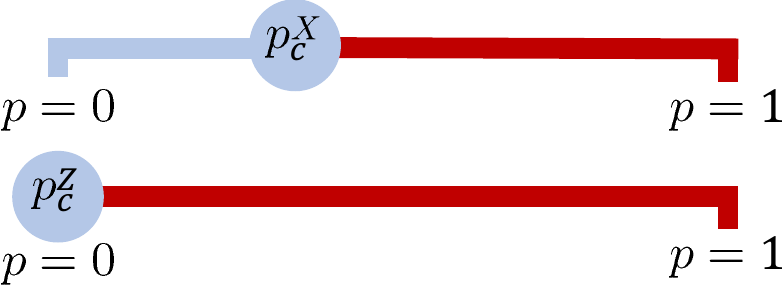}\label{fig:transition_diagram}}
    \hfill
    \sbox{\measurebox}{%
        \begin{minipage}[b]{0.5\columnwidth}
            \sidesubfloat[]{\includegraphics[width=0.8\columnwidth]{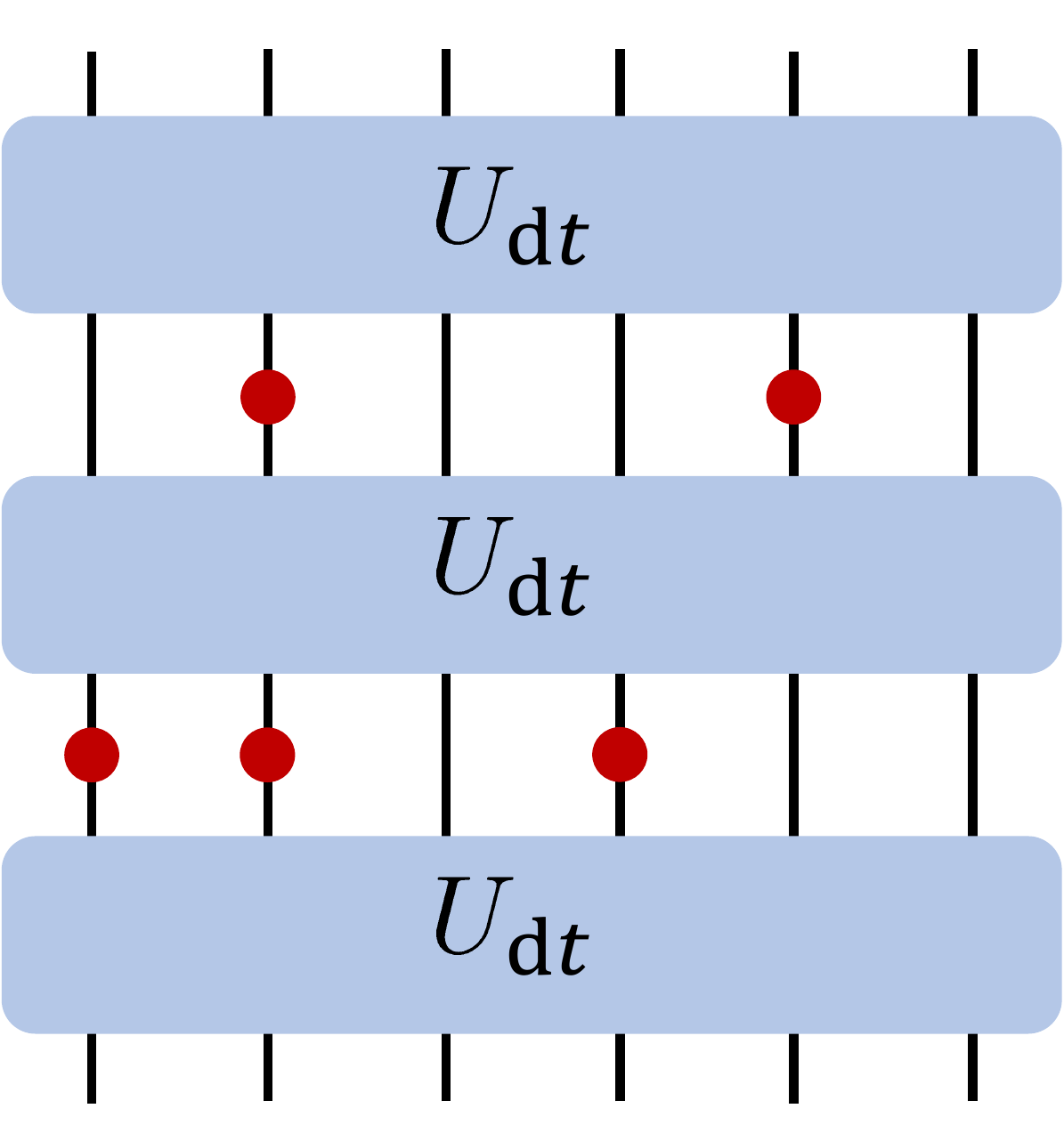}\label{fig:circuit_diagram}}
        \end{minipage}}
    \usebox{\measurebox}\hfill
    \begin{minipage}[b][\ht\measurebox][s]{0.5\columnwidth}
        \centering
        \sidesubfloat[]{\includegraphics[width=0.75\columnwidth]{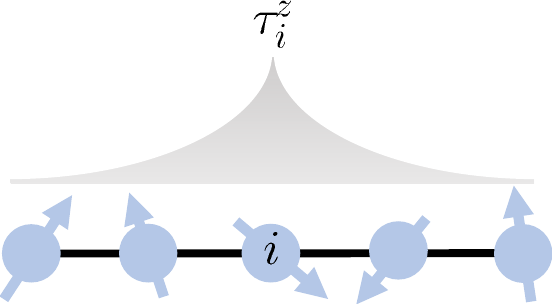}\label{fig:l_bit_diagram}}
        \vfill
        \sidesubfloat[]{\includegraphics[width=0.6\columnwidth]{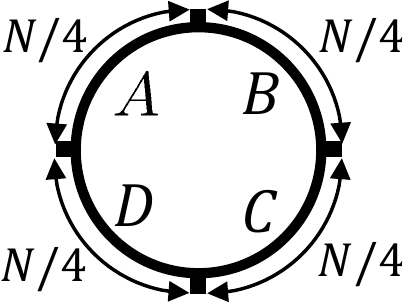}\label{fig:tripartite_info_diagram}}
    \end{minipage}
    \caption{\textbf{(a)} The critical properties of the transition depend on the measurement basis. Measurements in the X-basis result in a transition from volume-law to area-law entanglement at nonzero $p = p_{c}^{X} > 0$, similar to chaotic systems, whereas measurements in the Z-basis result in area-law entanglement for any nonzero $p$, similar to integrable systems. \textbf{(b)} Each stage of the dynamics consists of two steps. First the system is evolved in time with the unitary $U_{\diff t} = \exp(-i H \diff t)$, with $H$ given in \cref{eq:MBL_Hamiltonian}, and then for each spin we projectively measure in a tensor-product basis with probability $p$ (measurements are represented by the red dots). \textbf{(c)} A schematic of the typical support of an `l-bit' operator, localized at site $i$, in the fully MBL phase. \textbf{(d)} A depiction of the geometry used to calculate the tripartite information $I_{3}(A : B : C)$.}
\end{figure}

In this work we study the dynamics of an MBL Hamiltonian subjected to projective measurements. Many-body localized systems are characterized by a complete set of quasi-local operators, also known as `l-bits', which are conserved under the dynamics of the system's Hamiltonian (see \cref{fig:l_bit_diagram}) \cite{ husePhenomenologyFullyManybodylocalized2014, serbynUniversalSlowGrowth2013, imbrie2016many, rosIntegralsMotionManybody2015}. Due to this robust integrability, one might guess that the measurement-induced entanglement transition in MBL systems is similar to that in integrable systems. Although measurements in MBL and disorder-free integrable models share certain similarities, here we show that measurement-induced transitions performed in certain \textit{bases} (see \cref{fig:transition_diagram}) can have distinct properties.  

\floatsetup[figure]{style=plain,subcapbesideposition=top}

To illustrate this point, recall that in a chaotic quantum system local operators spread ballistically, resulting in the rapid scrambling of information with time which can be diagnosed through the decay of out-of-time-order correlators (OTOCs). The implication of this for the measurement-induced entanglement transition is that in chaotic systems the choice of measurement basis does not matter, provided it is local (i.e.\ close to a tensor product basis).
However, in an MBL system, only operators without any overlap with the l-bits are totally scrambled with their out-of-time ordered correlators decaying to zero. For example, the late-time limit of the disorder-averaged OTOC of an operator is set by its overlap with the local integrals of motion of the MBL system \cite{huang2017OTOC_MBL,PhysRevB.95.054201_RQHe, PhysRevB.95.060201_Swingle, chenOutoftimeorderCorrelationsManybody2017,  fan2017OTOC_MBL, maccormack2020OperatorgrowthMBL}. For the model described by \cref{eq:MBL_Hamiltonian}, in the strong-disorder limit $W \gg 1$ the l-bits $\{\tau_{i}^{z}\}$ are close to the local spin operators $\{S_{i}^{z}\}$, dressed by exponentially decaying tails \cite{chandranConstructingLocalIntegrals2015, Rademaker2016LIOM,  PhysRevLett.119.075701_Pekker,  PhysRevB.98.184201_Kulshreshtha, PhysRevB.97.134202_Goihl}, as shown in \cref{fig:l_bit_diagram}. The l-bits are related to the physical spins by a quasi-local unitary $U$ via $\tau_{i}^{z} = U S_{i}^{z} U^{\dag}$, with $U \to \mathds{1}$ in the strong-disorder limit. This means that the overlap of an operator $\mathcal{O}$ with an l-bit $\tau_{i}^{z}$ is given by $  \tr{\mathcal{O}\tau_{i}^{z}} = \tr{\mathcal{O} S_{i}^{z}} + \cdots$, where the dots indicate terms which vanish in the limit $W \to \infty$. As a result, the operator $S_{i}^{x}$ is scrambled by the MBL system, whereas the operator $S_{i}^{z}$ remains approximately localized.

This has the consequence that there is a qualitative difference between the nature of the measurement-induced entanglement transition in an MBL system depending on whether the measurements are performed in the basis of $S_{i}^{x}$ eigenstates (X-basis) or the basis of $S_{i}^{z}$ eigenstates (Z-basis), as shown in \cref{fig:transition_diagram}. With measurements in the X-basis, the transition from volume to area-law entanglement occurs at a nonzero measurement probability $p_{c}^{X} > 0$, similar to previously studied chaotic systems \cite{liQuantumZenoEffect2018,skinnerMeasurementInducedPhaseTransitions2019,baoTheoryPhaseTransition2020,chanUnitaryprojectiveEntanglementDynamics2019,jianMeasurementinducedCriticalityRandom2020,szyniszewskiEntanglementTransitionVariablestrength2019,gullansDynamicalPurificationPhase2019,gullansScalableProbesMeasurementinduced2019,liMeasurementdrivenEntanglementTransition2019,choiQuantumErrorCorrection2019,zabaloCriticalPropertiesMeasurementinduced2020,tangMeasurementinducedPhaseTransition2020,gotoMeasurementInducedTransitionsEntanglement2020,liConformalInvarianceQuantum2020,lopez-piqueresMeanfieldTheoryEntanglement2020,fanSelfOrganizedErrorCorrection2020,zhangNonuniversalEntanglementLevel2020}. On the other hand, with measurements in the Z-basis, the volume-law is destroyed for any nonzero $p$, similar to previously studied integrable systems \cite{caoEntanglementFermionChain2019,rossiniMeasurementinducedDynamicsManybody2020,chanUnitaryprojectiveEntanglementDynamics2019}.  

This difference can also be related to the interplay between measurements and the phenomenology of entanglement growth in MBL systems. As we discuss in \cref{sec:diagonal_entropy}, in a measurement-free MBL system, the steady state entanglement is governed by the diagonal entropy, which is constant under Hamiltonian time evolution. Here we show that measurements induce dynamics in the diagonal entropy which qualitatively differ between X-basis measurements and Z-basis measurements, where X-basis measurements tend to maximize the diagonal entropy while Z-basis measurements tend to minimize it. This then drives the steady state to volume-law or area-law entanglement, depending on the measurement basis.

In a recent work \cite{ippolitiEntanglementPhaseTransitions2020}, the authors investigated the dynamics of a finite-range l-bit model of MBL under local Clifford dynamics. We analyze the full l-bit Hamiltonian, which includes infinite-range interactions between l-bits. It also provides a picture of the measurement-induced entanglement transition in a fully realistic model of MBL and our results are consistent with results in Ref.\ \cite{ippolitiEntanglementPhaseTransitions2020} where they can be compared.

\section{Model}
\label{sec:model}

We study a standard model of MBL, the disordered spin-$\frac{1}{2}$ Heisenberg chain, with a Hamiltonian given by

\begin{equation}
    H = J \sum_{\langle i j \rangle} \mathbf{S}_{i} \cdot \mathbf{S}_{j} + \sum_{i=1}^{N} h_{i} S_{i}^{z} 
    \label{eq:MBL_Hamiltonian}
\end{equation}

where the $h_{i}$ are random variables drawn from the uniform distribution on $[-h,h]$, and the double sum runs over nearest neighbours. We use periodic boundary conditions unless stated otherwise. For sufficiently large disorder strength $W \equiv h/J$ this model is many-body localized. Early studies of MBL based on exact diagonalization estimated the critical disorder strength $W_{c}$ to be around $W_{c} \approx 3.6$ \cite{palManybodyLocalizationPhase2010, luitz2015many}, though more recent studies have argued that $W_{c}$ has significant finite-size corrections, and have suggested a figure of $W_{c} \approx 5$ \cite{doggenManybodyLocalizationDelocalization2018}. In this paper we are interested in the region well into the localized phase, and so will take $W = 10$, large enough to avoid issues with finite-size drifts of $W_{c}$.

To study the effects of measurements on the entangling properties of MBL dynamics, we consider a repeated two-stage protocol, as illustrated in \cref{fig:circuit_diagram}. Starting from an initial random product state, we apply unitary dynamics generated by the Hamiltonian for some time $\diff t$. Unless otherwise specified we take $\diff t = 1$ in units of $J$, independent of $N$, to make the situation comparable with random local circuit models. Then for each spin we projectively measure in a local tensor-product basis with probability $p$, where $p$ can be interpreted as the density of measurements in spacetime. This process is repeated until the entanglement entropies reach a steady state.

\subsection{Transition diagnostics}

To study the measurement-induced entanglement transition in this MBL system, we performed the time-evolution using exact diagonalization. We made use of the fact that the Hamiltonian $H$ in \cref{eq:MBL_Hamiltonian} conserves total $S^{z}$, so in a basis ordered by total $S^{z}$, $H$ is block-diagonal and one can diagonalize the blocks separately.

To characterize the transition we focused on two quantities: the von Neumann entropy $S_{X} = -\tr{\rho_{X} \ln{\rho_{X}}}$, and the tripartite information

\begin{equation}
        I_{3}(A : B : C) = I(A : B) + I(A : C) - I(A : BC),
\end{equation}

where $I(A : B) = S_{A} + S_{B} - S_{AB}$ is the mutual information. We calculate $I_{3}(A : B : C) \equiv I_{3}$ for the geometry shown in \cref{fig:tripartite_info_diagram}. It is easy to show that for a system partitioned into four subsystems, the tripartite information of any three of the subsystems does not depend on the choice of subsystems, so once the partitioning is fixed there is no ambiguity in calling this quantity $I_{3}$.

As discussed in Refs.\ \cite{zabaloCriticalPropertiesMeasurementinduced2020,gullansDynamicalPurificationPhase2019}, the advantage of using the tripartite information here is that it avoids any $\log{N}$ divergences in the entanglement entropy at criticality, which have been observed in hybrid Haar-random circuit models \cite{zabaloCriticalPropertiesMeasurementinduced2020}. Instead, $I_{3}$ is expected to scale as $I_{3} \propto -N$ in the volume-law phase, reach an $\mathcal{O}(1)$ constant at criticality, and then vanish in the area-law phase. This is especially important given the limited system sizes accessible by exact diagonalization.

Throughout this paper we will mostly focus on the von Neumann entropy, and quantities defined in terms of it. However, one can also consider the transitions in the broader family of R\'{e}nyi entropies, defined as

\begin{equation}
    S_{n}(\rho) = \dfrac{1}{1-n}\ln\mathrm{Tr}\left[\rho^{n}\right].
\end{equation}

The R\'{e}nyi-$n$ entropy tends to the von Neumann entropy in the limit $n \to 1$. For $n > 1$, there is the inequality $S_{\infty} \leq S_{n} \leq \frac{n}{n-1} S_{\infty}$, which implies that all R\'{e}nyi-$n$ entropies with $n>1$ must have the same scaling with system size. We consider the transition in the R\'{e}nyi entropies in \cref{sec:X_basis_measurements}.

\begin{figure}[t]
    \begin{center}
        \includegraphics[width=\columnwidth]{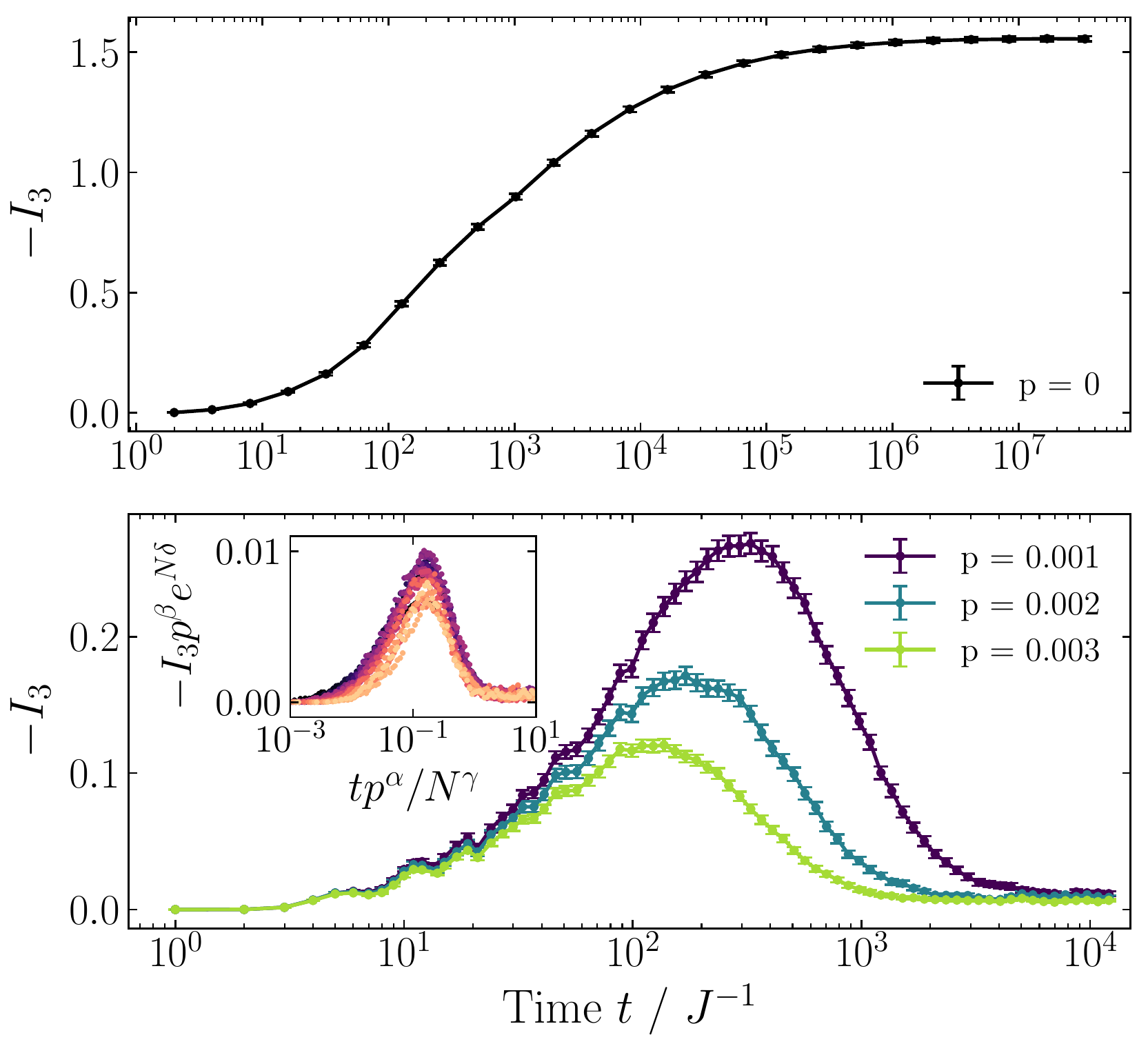}
    \end{center}
    \caption{A comparison of the dynamics of the tripartite information $I_{3}$ for $p=0$ and $p>0$ with Z-basis measurements, with $N=12$. For $p=0$, $I_{3}$ grows logarithmically in time before saturating to a volume-law. For $p>0$, $I_{3}$ initially grows in time, reaching an area-law peak, before decaying to a constant which decreases with system size. The inset shows that the dynamics can be reasonably well described by the scaling form $-I_{3}(t,p,N) = F\left[t p^{\alpha}/N^{\gamma}\right] / p^{\beta} e^{N\delta}$, where $F$ is a scaling function and $\alpha = 0.80$, $\beta = 0.78$, $\gamma = 0.81$ and $\delta = 0.16$. The inset data corresponds to 20 separate time series, with $N=12,16$ and $0.0005 \leq p \leq 0.005$.}
    \label{fig:Z_basis_vN_I3_dynamics}
\end{figure}
\begin{figure}[t]
    \centering
    \subfloat[][]{
        \centering
        \includegraphics[width=0.475\columnwidth]{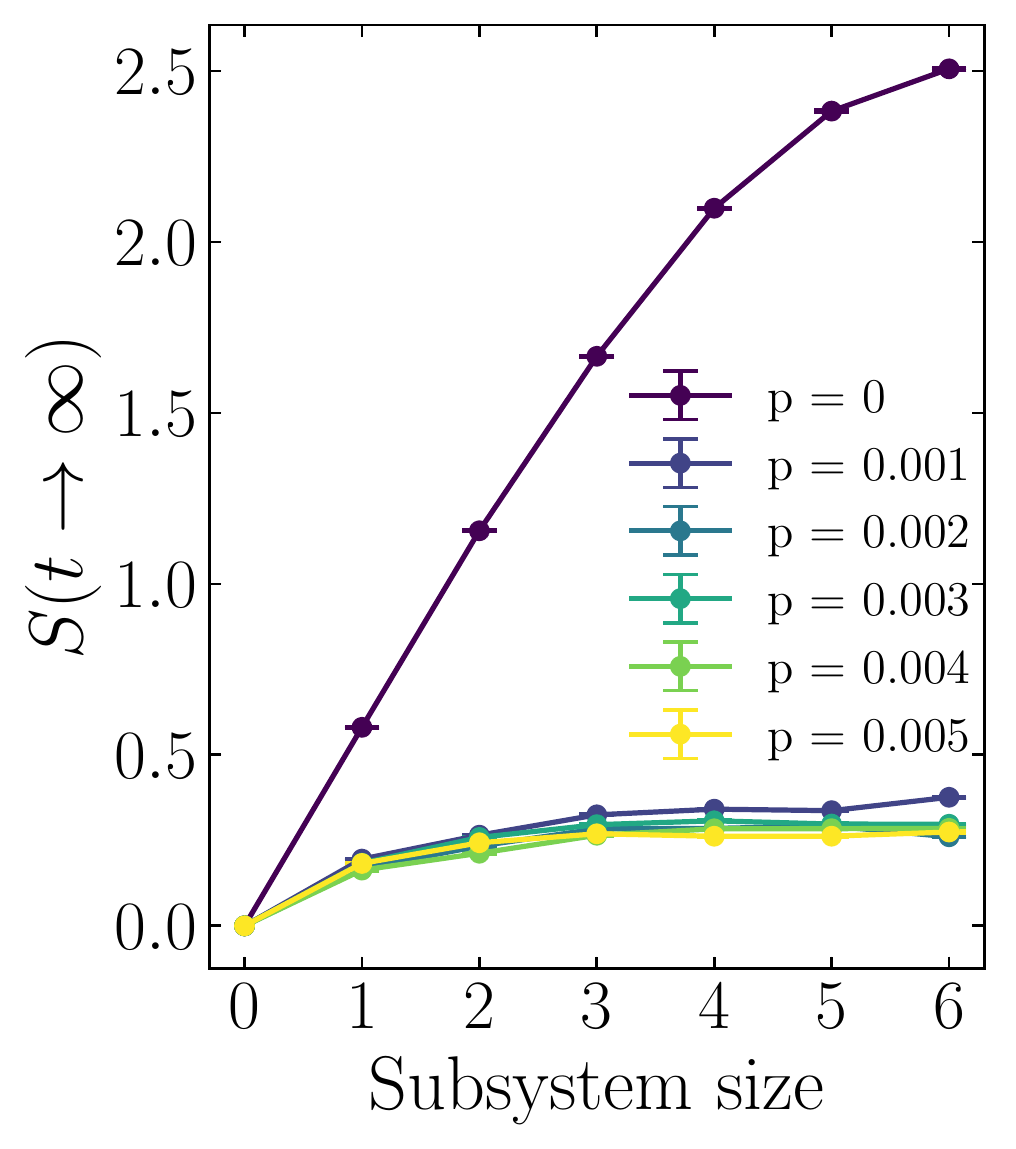}
        \label{fig:saturation_vn_entropy_vs_subsystem_size_Z_basis_measurements}}
    \subfloat[][]{
        \centering
        \includegraphics[width=0.5\columnwidth]{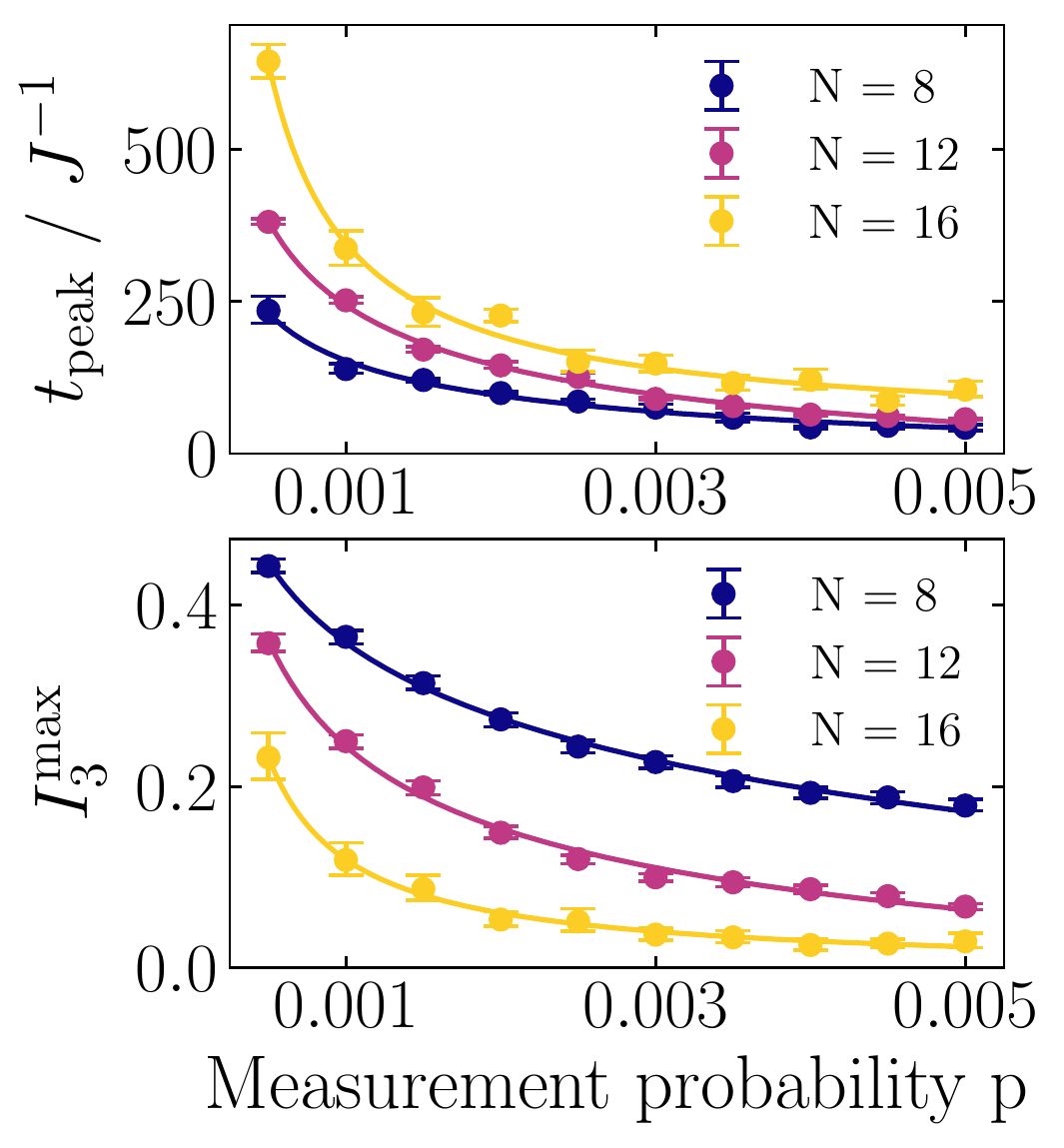}
        \label{fig:peak_finite_size_scaling}}
    \caption{\textbf{(a)} Scaling of the von Neumann entropy at saturation with subsystem size for a system of size $N=12$. For $p=0$ this quantity obeys a volume-law, but for $p>0$ it saturates to an area-law. \textbf{(b)} Scaling of the time $t_{\mathrm{peak}}$ and height $I_{3}^{\mathrm{max}}$ of the intermediate time peak in the tripartite information $I_{3}$, as seen in \cref{fig:Z_basis_vN_I3_dynamics}. The solid lines are power law fits, decaying approximately as $p^{-1}$. With increasing system size, the peak occurs at later times and is smaller in magnitude, indicating that it is (sub-)area-law.} 
\end{figure}

\subsection{Diagonal entropy}
\label{sec:diagonal_entropy}

The qualitative difference in the dynamics of the entanglement entropies with Z- or X-basis measurements can also be related to the interplay of measurements with the phenomenology of entanglement growth in MBL systems. In the absence of measurements, the steady-state entanglement of an MBL system is set by the diagonal entropy of the initial state in the energy eigenbasis \cite{znidaricManybodyLocalizationHeisenberg2008,serbynUniversalSlowGrowth2013,bardarsonUnboundedGrowthEntanglement2012,husePhenomenologyFullyManybodylocalized2014,nanduriEntanglementSpreadingManybody2014}, $S(t \to \infty) \propto \sdiag \left( |\psi(0) \rangle \right)$, with

\begin{equation}
    \sdiag \left( |\psi(0) \rangle \right) = - \sum_{i} p_{i} \ln{p_{i}}, \quad p_{i} = \left| \langle E_{i} | \psi(0) \rangle \right|^{2},
\end{equation}

where the $\{|E_{i}\rangle\}$ are the eigenstates of $H$. This is a result of dephasing between l-bits. For many classes of initial states, such as random product states, the average diagonal entropy is extensive in the strong disorder regime, thus giving rise to volume-law steady state entanglement in MBL systems. 

It is worth noting that $\sdiag$ is constant under time-evolution by $H$, since it depends only on the magnitude of the amplitudes $\langle E_{i} | \psi(t) \rangle$, which merely pick up a phase. However, once measurements are introduced, the diagonal entropy is no longer preserved in time. Instead, the measurements drive the diagonal entropy to a new steady state, typically at an exponential rate $\lambda \approx p$. The choice of measurement basis determines the nature of this new steady state. In the strong disorder limit, the eigenstates of $H$ are close to product states of $S_{i}^{z}$ eigenstates. Thus measurements in the Z-basis tend to drive the diagonal entropy close to zero (see \cref{fig:diagonal_entropy_dynamics_Z_basis_measurements}), whereas measurements in the X-basis tend to drive it close to its maximum value (see \cref{fig:diagonal_entropy_dynamics_X_basis_measurements}).

\section{Entanglement dynamics with Z-basis measurements}
\label{sec:Z_basis_measurements}

In this section we focus on unitary MBL dynamics interspersed with random projective measurements in the Z-basis, and demonstrate the fact that there is a qualitative difference once measurements are introduced with any arbitrarily small probability $p$. That this transition occurs at $p_{c}^{Z}=0$ is similar to previously studied integrable systems \cite{caoEntanglementFermionChain2019,rossiniMeasurementinducedDynamicsManybody2020,chanUnitaryprojectiveEntanglementDynamics2019}, which is consistent with the picture of MBL as a form of quasi-integrability  robust to local perturbations.

\begin{figure}[b]
    \begin{center}
        \includegraphics[width=\columnwidth]{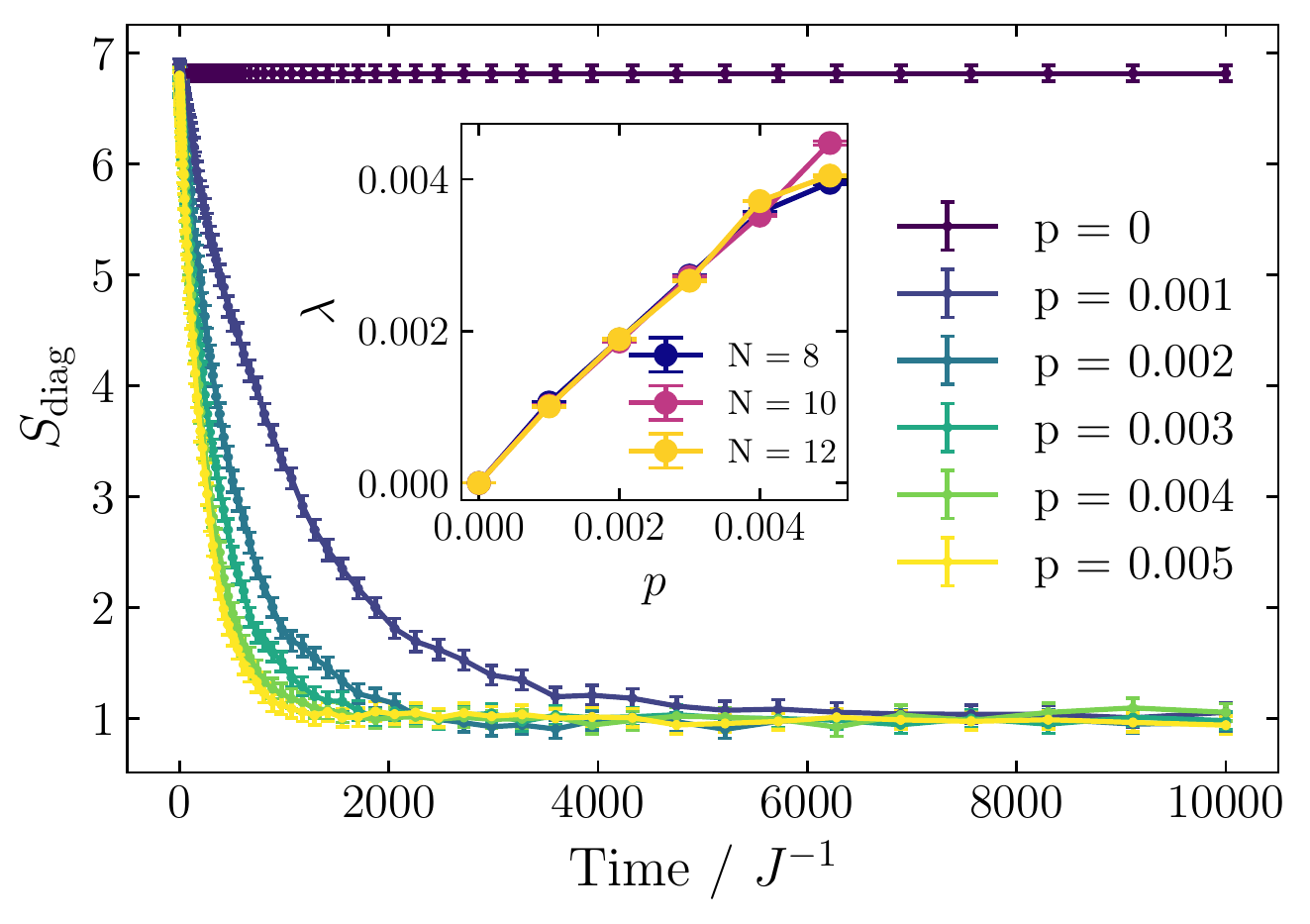}
    \end{center}
    \caption{Dynamics of the diagonal entropy $\sdiag$ using the energy eigenbasis, with $N=12$. $\sdiag$ is preserved by Hamiltonian evolution, so for $p=0$ it remains constant in time. For $p>0$, projective Z-basis measurements cause it to decay exponentially at a rate proportional to $p$. The inset shows the dependence of the early-time decay rate $\lambda$ on $p$ and $N$, with $\lambda \approx p$, independent of $N$.}
    \label{fig:diagonal_entropy_dynamics_Z_basis_measurements}
\end{figure}

\cref{fig:Z_basis_vN_I3_dynamics} shows a comparison of the dynamics of the tripartite information $I_{3}$ between $p=0$ and $p>0$. For $p=0$, $I_{3}$ grows in magnitude logarithmically in time before saturating to a volume-law, similar to the well-known behaviour of the half-chain entanglement entropy. On the other hand, for $p>0$ the tripartite information initially grows at early times while the diagonal entropy is still large, reaching an area-law peak (see \cref{fig:peak_finite_size_scaling}), but eventually the measurements dominate and $I_{3}$ decays to a smaller area-law. This area-law peak also appears in the Bell pair model of Ref.\ \cite{chanUnitaryprojectiveEntanglementDynamics2019}, but there it occurs at an $\mathcal{O}(1)$ time $t_{\mathrm{peak}} \sim 1/p$. In this system, the peak time does still scale as $1/p$, but it also increases with system size (see top panel of \cref{fig:peak_finite_size_scaling}). We expect that if measurements were performed directly in the l-bit basis, then this peak would be absent. The fact that our choice of basis differs from the l-bit basis in its lack of exponential tails may give rise to this intermediate-time peak.

The inset to \cref{fig:Z_basis_vN_I3_dynamics} shows that the dynamics can be reasonably well described by the scaling form $-I_{3}(t,p,N) = F\left[t p^{\alpha}/N^{\gamma}\right] / p^{\beta} e^{N\delta}$, where $F$ is a scaling function and $\alpha = 0.80$, $\beta = 0.78$, $\gamma = 0.81$ and $\delta = 0.16$. Hence in the thermodynamic limit $N\to \infty$ we expect the steady-state tripartite information to scale rapidly to zero with $N$ for any $p>0$.

Focusing further on the transition in the steady state, \cref{fig:saturation_vn_entropy_vs_subsystem_size_Z_basis_measurements} shows the von Neumann entropy of contiguous subsystems of different sizes in the steady state. Whereas for $p=0$ this quantity is extensive, for $p>0$ it quickly becomes independent of the size of the subsystem, indicating an area-law. As discussed in \cref{sec:introduction}, this behaviour can be linked to the measurement-induced dynamics of the diagonal entropy $\sdiag$, which governs the steady state entanglement in the measurement-free system. \cref{fig:diagonal_entropy_dynamics_Z_basis_measurements} shows that, for any $p>0$, the diagonal entropy decays exponentially in time, before reaching a steady state which is independent of $p$. The inset to \cref{fig:diagonal_entropy_dynamics_Z_basis_measurements} shows that the decay rate $\lambda$ scales approximately as $\lambda \approx p$, independent of system size. We expect this result to hold throughout the fully MBL regime as a result of the strongly localized nature of the l-bits, and the fact that the measurements are performed independently on each site.

\section{Entanglement dynamics with X-basis measurements}
\label{sec:X_basis_measurements}

\begin{figure}[t]
    \begin{center}
        \includegraphics[width=\columnwidth]{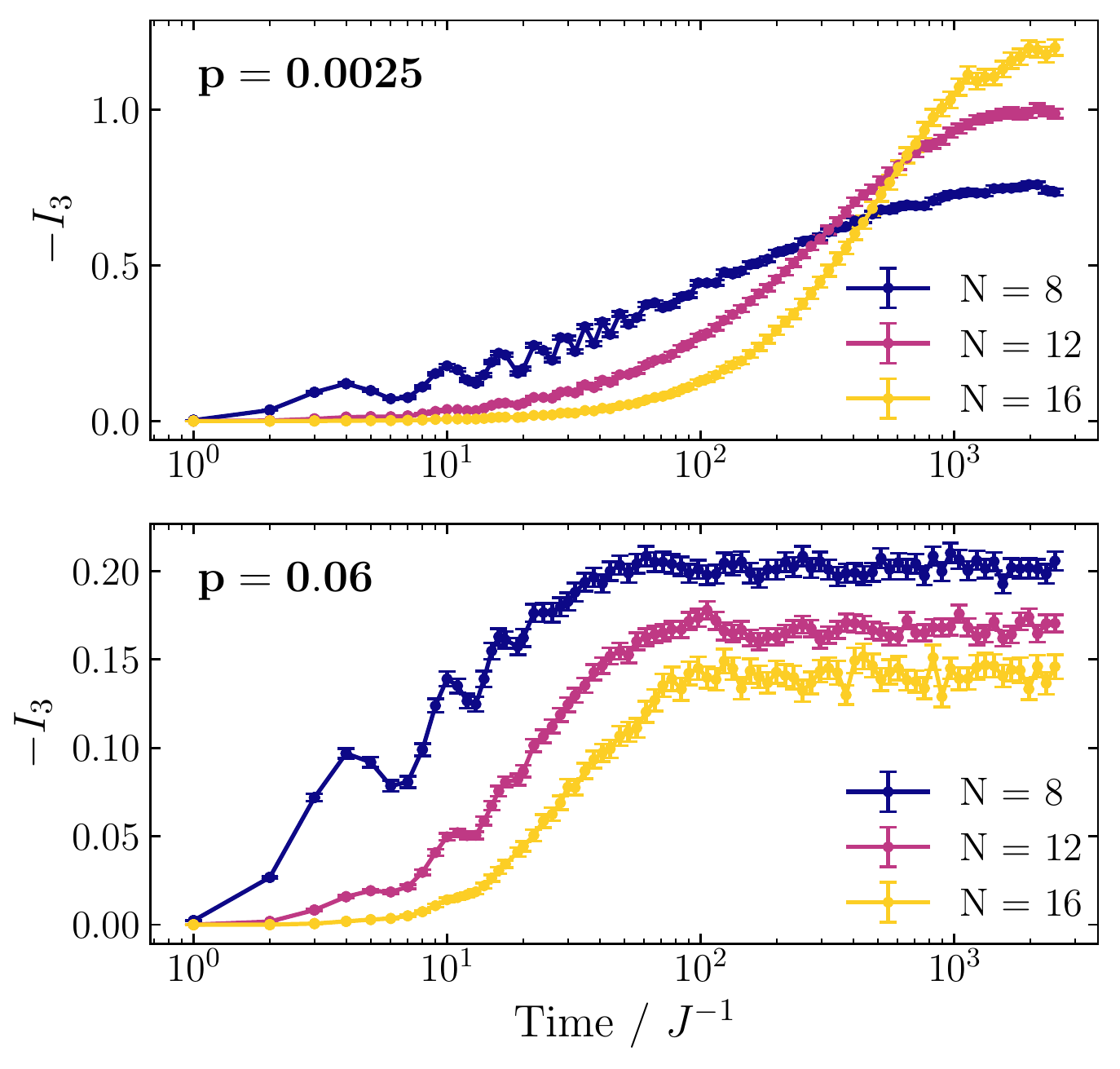}
    \end{center}
    \caption{A comparison of the dynamics of the tripartite information $I_{3}$ with X-basis measurements for values of $p$ below and above the transition at $p_{c}^{X} \approx 0.014$. For $p=0.0025$, $I_{3}$ grows logarithmically in time before saturating to a volume-law. By $p=0.06$, the system size scaling has reversed, indicating a transition to an area law.}
    \label{fig:X_basis_vN_I3_dynamics}
\end{figure}

\begin{figure}[t]
    \begin{center}
        \includegraphics[width=\columnwidth]{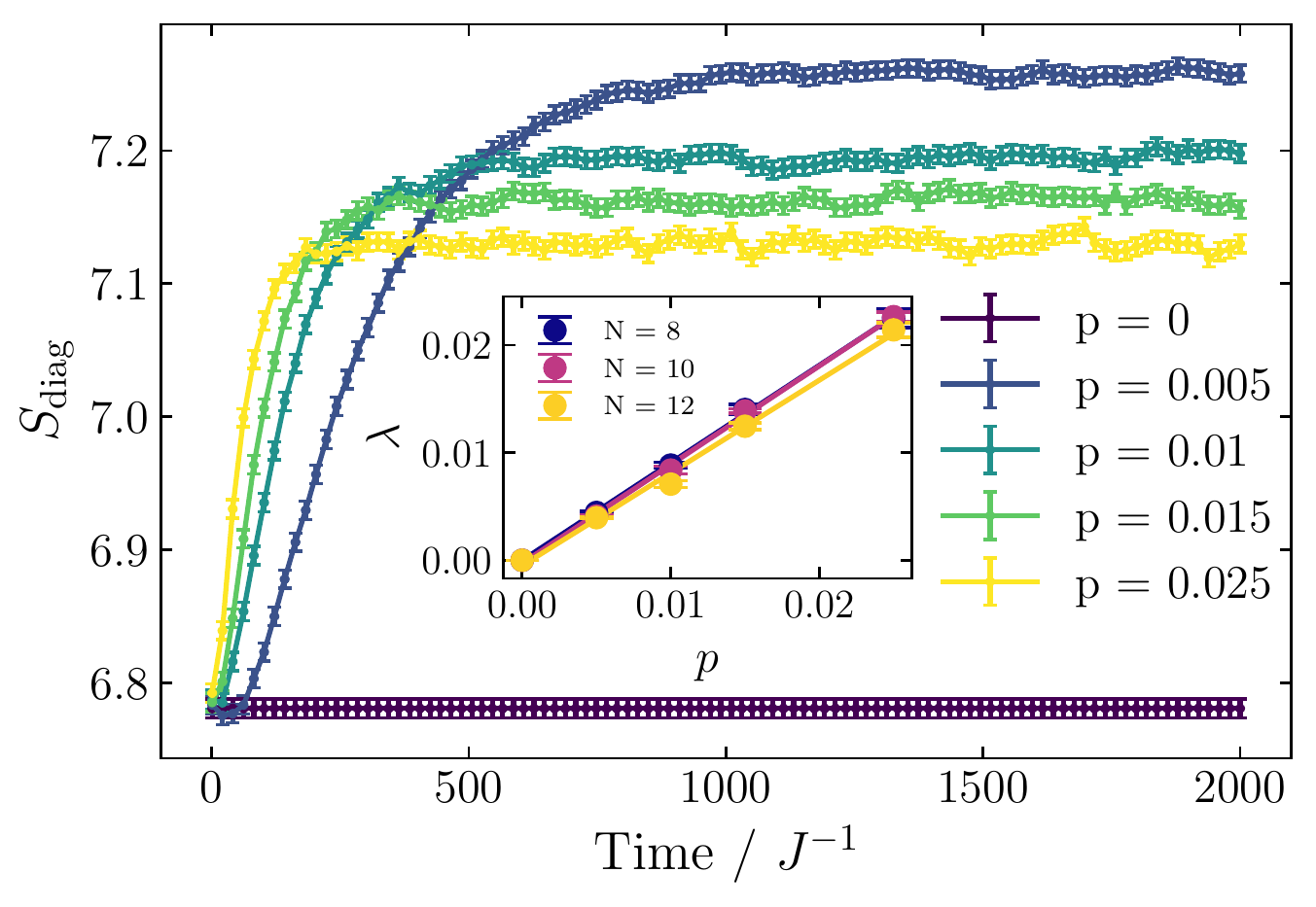}
    \end{center}
    \caption{Dynamics of the diagonal entropy $\sdiag$ using the energy eigenbasis, with $N=12$. $\sdiag$ is preserved by Hamiltonian evolution, so for $p=0$ it remains constant in time. For $p>0$, projective X-basis measurements cause it grow as $S_{\mathrm{diag}}(t) = S_{\mathrm{diag}}(0) e^{-\lambda t} + S_{\mathrm{diag}}(\infty) (1-e^{-\lambda t})$ at a rate $\lambda$ proportional to $p$. The inset shows the dependence of the early-time decay rate $\lambda$ on $p$ and $N$, with $\lambda \approx p$, independent of $N$.}
    \label{fig:diagonal_entropy_dynamics_X_basis_measurements}
\end{figure}

Having seen in \cref{sec:Z_basis_measurements} the collapse of the MBL steady-state volume-law when the dynamics are interspersed with arbitrarily rare Z-basis measurements, we now demonstrate that this volume-law is more resilient to measurements in the X-basis. More precisely, we aim to show that there is a \textit{nonzero} critical measurement probability $p_{c}^{X}>0$, below which the steady-state volume-law persists. This scenario is similar to the case in previously studied chaotic systems \cite{liQuantumZenoEffect2018,skinnerMeasurementInducedPhaseTransitions2019,baoTheoryPhaseTransition2020,chanUnitaryprojectiveEntanglementDynamics2019,jianMeasurementinducedCriticalityRandom2020,szyniszewskiEntanglementTransitionVariablestrength2019,gullansDynamicalPurificationPhase2019,gullansScalableProbesMeasurementinduced2019,liMeasurementdrivenEntanglementTransition2019,choiQuantumErrorCorrection2019,zabaloCriticalPropertiesMeasurementinduced2020,tangMeasurementinducedPhaseTransition2020,gotoMeasurementInducedTransitionsEntanglement2020,liConformalInvarianceQuantum2020,lopez-piqueresMeanfieldTheoryEntanglement2020,fanSelfOrganizedErrorCorrection2020,zhangNonuniversalEntanglementLevel2020}.

The top panel of \cref{fig:X_basis_vN_I3_dynamics} shows the dynamics of the tripartite information $I_{3}$ for nonzero $p = 0.0025$, which is below the critical measurement probability $p_{c}^{X} \approx 0.014$. Here the dynamics are qualitatively similar to those for $p = 0$, shown in the top panel of \cref{fig:Z_basis_vN_I3_dynamics}, where the magnitude of $I_{3}$ grows logarithmically in time before saturating to a steady-state value which increases with system size. The bottom panel of \cref{fig:X_basis_vN_I3_dynamics} shows that, above the transition $p > p_{c}^{X}$, the magnitude of $I_{3}$ still grows logarithmically in time, but no longer saturates to an extensive steady state, indicating a transition to an area law. The growth of the diagonal entropy is qualitatively similar for all nonzero $p$---it follows the functional form $S_{\mathrm{diag}}(t) = S_{\mathrm{diag}}(0) e^{-\lambda t} + S_{\mathrm{diag}}(\infty) (1-e^{-\lambda t})$, with the growth rate $\lambda$ proportional to $p$, as shown in \cref{fig:diagonal_entropy_dynamics_X_basis_measurements}. Interestingly, in contrast to the case with Z-basis measurements, here $\sdiag(\infty)$ decreases with $p$. This may be a consequence of the finite disorder-strength, which means that the energy eigenstates aren't exactly product states.

It is also worth noting that, as a result of the increase in diagonal entropy induced by X-basis measurements, it is possible to boost the steady state entanglement relative to $p=0$, contrary to the usual picture of measurements only destroying entanglement. Of course, this effect is relevant only for fairly small $p$, since for larger $p$ the disentangling power of the individual measurements overcomes the boost from the increased diagonal entropy. 

\begin{figure}[t]
    \begin{center}
        \includegraphics[width=\columnwidth]{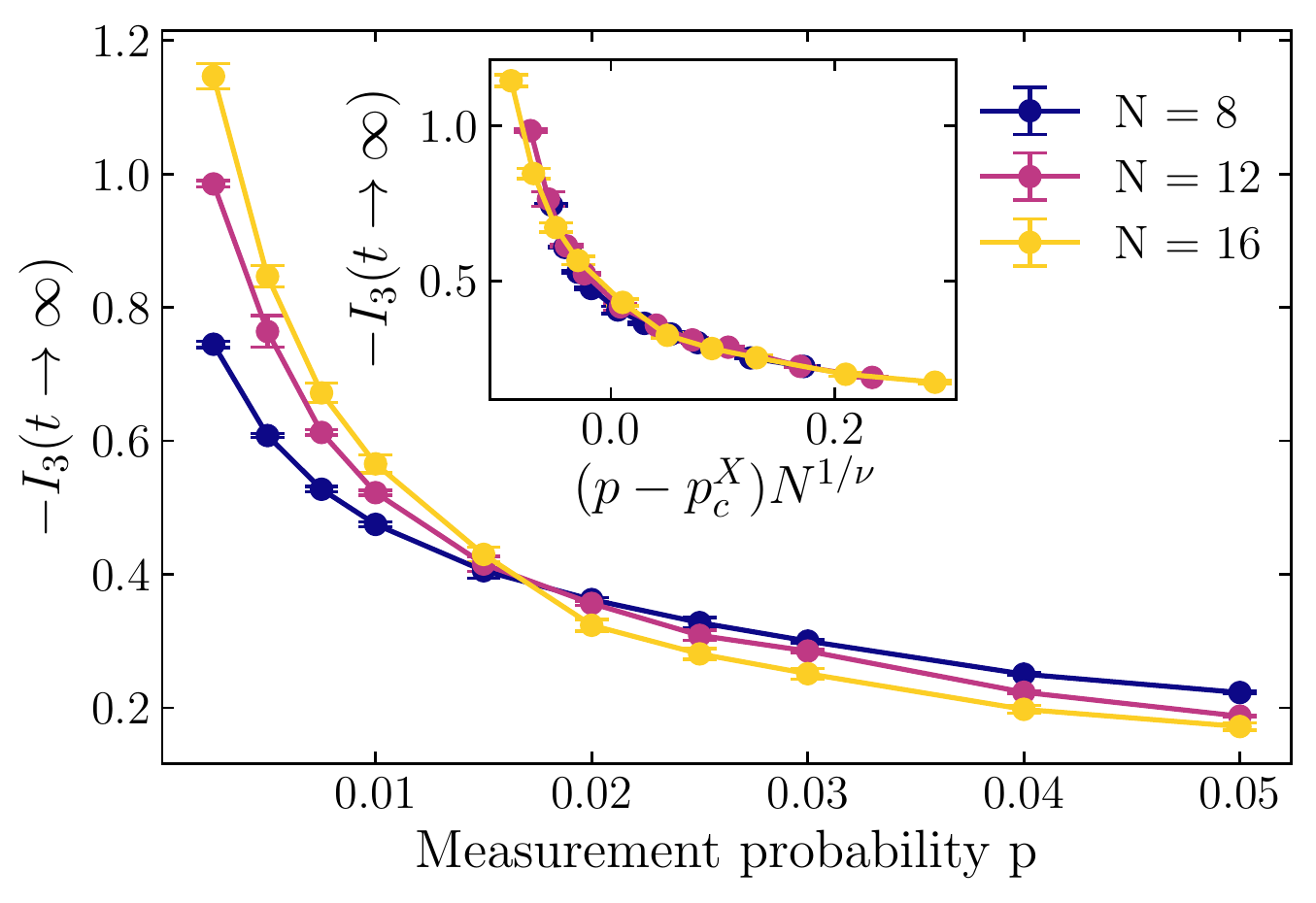}
    \end{center}
    \caption{The steady state tripartite information $I_{3}$ as a function of measurement probability $p$ and system size $N$. The inset shows a data collapse with the fitted parameters $p_{c}^{X} = 0.014(2)$ and $\nu = 1.3(2)$.}
    \label{fig:X_basis_vN_I3_steady_state}
\end{figure}

\begin{figure}[t]
    \centering
    \includegraphics[width=0.9\columnwidth]{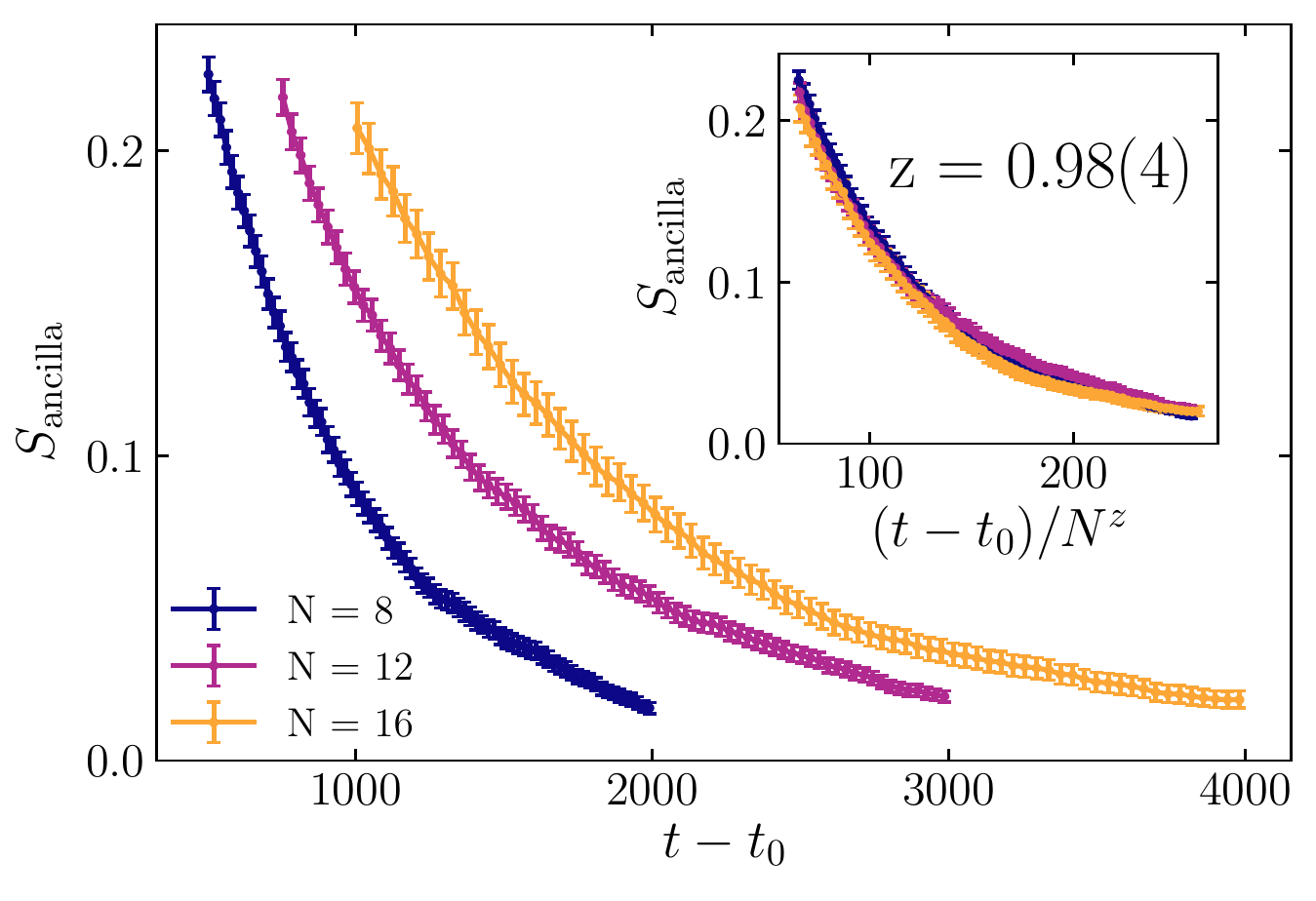}
    \caption{Extracting the dynamical critical exponent $z=0.98(4)$ using the entanglement entropy of an ancilla qubit as an order parameter for the transition \cite{gullansScalableProbesMeasurementinduced2019,zabaloCriticalPropertiesMeasurementinduced2020}. The ancilla is maximally entangled with the bulk system at the saturation time $t_{0}$ determined by the dynamics of $I_{3}$, and then we continue with the dynamics described in \cref{sec:model} (acting only on the original spins). In performing the data collapse, we exclude times shortly after $t_{0}$ because the scaling only occurs after an intermediate timescale.}
    \label{fig:dynamical_exponent}
\end{figure}

With the goal of studying the transition point between the volume- and area-law phases, we show in \cref{fig:X_basis_vN_I3_steady_state} the steady state tripartite information $I_{3}$ as a function of measurement probability $p$ and system size $N$. To estimate the properties of the critical point, we assume a scaling function of the form $I_{3}(t \to \infty, p, N) = F\left[(p-p_{c}^{X}) N^{1/\nu}\right]$, and perform a fit to minimize the least-squares distance between each scaled point and the line obtained by a linear interpolation between its neighbours (see the supplementary material of Ref.\ \cite{zabaloCriticalPropertiesMeasurementinduced2020} for more details). This yields the parameters $p_{c}^{X} = 0.014(2)$ and $\nu = 1.3(2)$, where the error bars correspond to the region where the cost function from the fit is less than 1.3 times its minimum. We expect $p_{c}^{X}$ to vary with the time step $\diff t$, so we do not believe it will display universal behaviour. However, the critical exponent $\nu$ is close to the value of $\nu  = \frac{4}{3}$ for 2+0D percolation, similar to the results in random local unitary circuits \cite{zabaloCriticalPropertiesMeasurementinduced2020}. To test for the presence of conformal symmetry at the critical point, we also extract the dynamical critical exponent $z$ using the method described in Refs.\ \cite{gullansScalableProbesMeasurementinduced2019,zabaloCriticalPropertiesMeasurementinduced2020} of using the entanglement entropy of a single ancilla qubit as an order parameter for the transition. This yields $z = 0.98(4)$, as shown in \cref{fig:dynamical_exponent}, close to the value of $z=1$ for conformal symmetry. 

\floatsetup[figure]{style=plain}
\begin{figure}[t]
    \centering
    \sidesubfloat[]{\includegraphics[width=0.9\columnwidth]{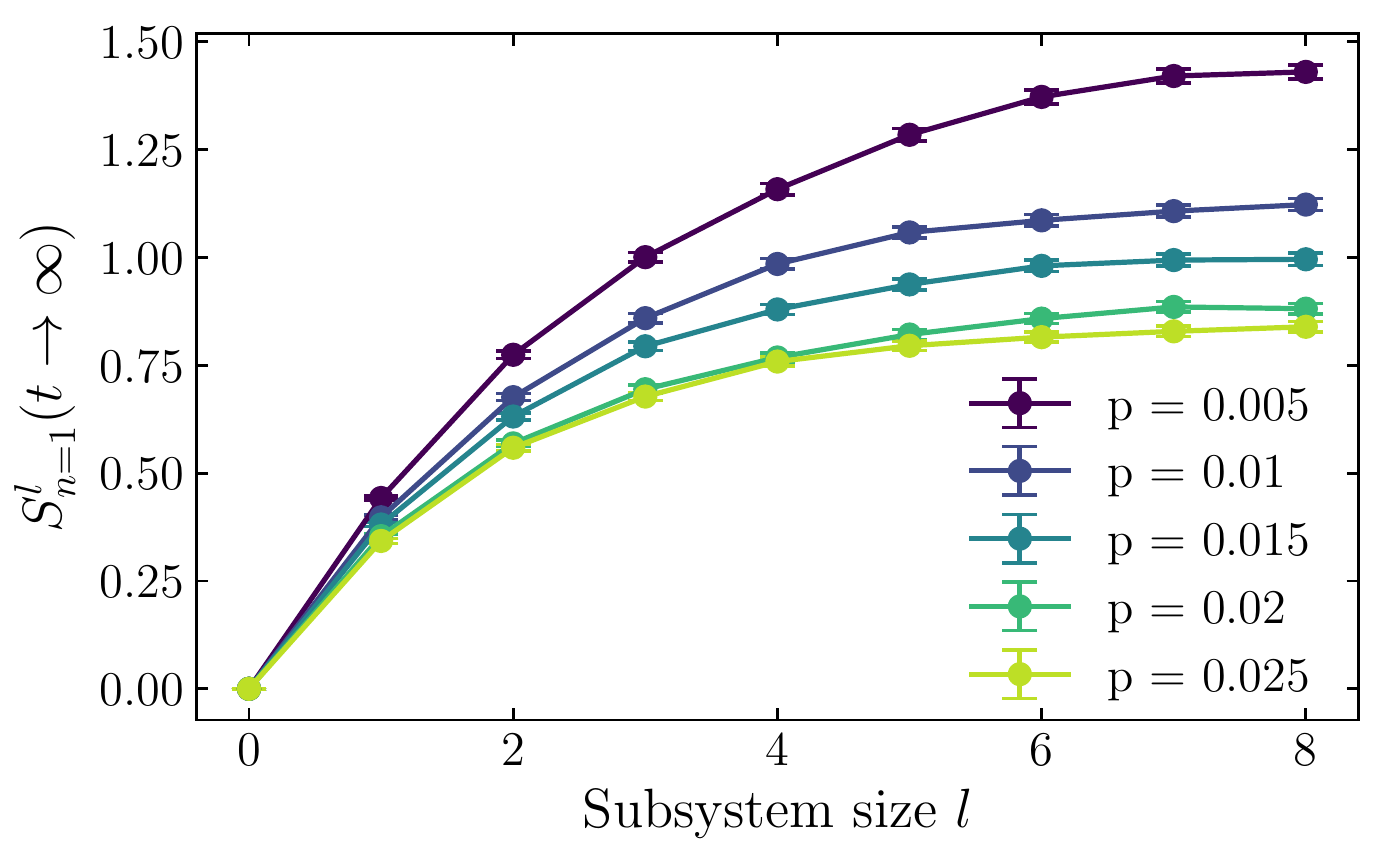}\label{fig:X_basis_steady_state_vn}}
    
    \sidesubfloat[]{\hspace{4pt}\includegraphics[width=0.42\columnwidth]{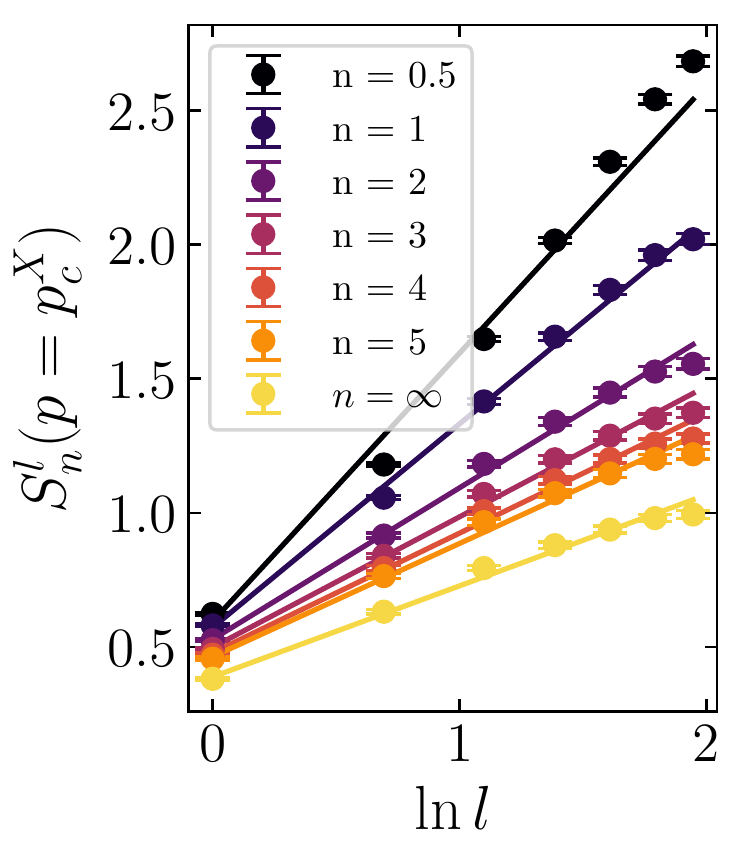}\label{fig:X_basis_steady_state_vn_log_scaling}}
    \sidesubfloat[]{\hspace{-4pt}\includegraphics[width=0.42\columnwidth]{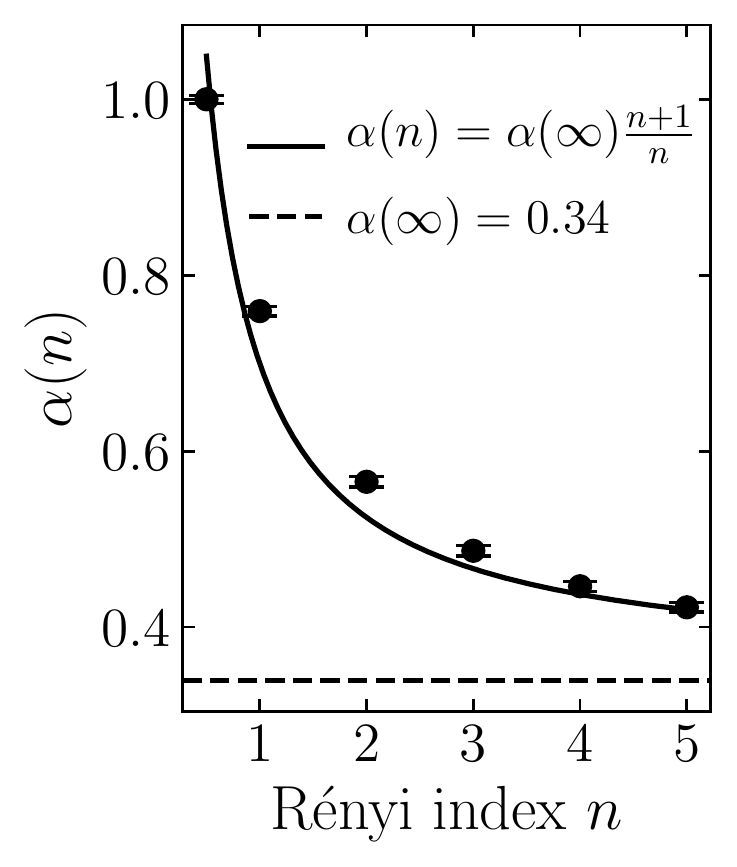}\label{fig:X_basis_steady_state_vn_log_scaling_coefficients}}
    \caption{\textbf{(a)} Scaling of the steady-state von Neumann entropy with subsystem size $l$. The transition from volume to area-law occurs at $p_{c}^{X} \approx 0.014$. \textbf{(b)} The steady-state R\'{e}nyi-$n$ entropies at criticality as a function of $\ln{l}$. The straight lines are fits to $S_{n}^{l}(p=p_{c}^{X}) = \alpha(n) \ln{l} + \beta(n)$. \textbf{(c)} The dependence of the log coefficient $\alpha(n)$ on the R\'{e}nyi index $n$. The solid line is a fit to the form $\alpha(n) = \alpha(\infty)(1 + 1/n)$, where $\alpha(\infty) \approx 0.34$.}
\end{figure}

Finally, having obtained an estimate for the critical point, we examine the scaling of the steady-state entanglement entropy $S^{l}_{n}(p=p_{c}^{X})$ with subsystem size $l$ at criticality, where $n$ indicates the R\'{e}nyi index. In random circuit models with interspersed measurements, $S^{l}_{n}(p=p_{c}^{X})$ was found to scale logarithmically with $l$, suggesting an underlying conformal field theory (CFT) description. In \cref{fig:X_basis_steady_state_vn_log_scaling}, we plot $S^{l}_{n}(p=p_{c}^{X})$ as a function of $\ln{l}$, and find a similar logarithmic scaling, albeit for fairly small subsystem sizes. The coefficient $\alpha(n)$ of the log term depends on the R\'{e}nyi index $n$, as shown in \cref{fig:X_basis_steady_state_vn_log_scaling_coefficients}, with the dependence well described by $\alpha(n) = \alpha(\infty) (1 + 1/n)$, which is the behaviour expected for unitary CFTs \cite{calabreseEntanglementSpectrumOnedimensional2008}. Interestingly, in Ref.\ \cite{zabaloCriticalPropertiesMeasurementinduced2020} the authors found for spin-$\frac{1}{2}$ Haar-random circuits that $\alpha(n)$ was not quite described by a fit of this form, but rather required an offset, $\alpha(n) = a(1+1/n) + b$. This departure from the behaviour of unitary CFTs could be seen as consistent with the fact that, in the analytically solvable limit of infinite local Hilbert space dimension $q \to \infty$, the measurement-induced transition in a Haar-random circuit is described by a \textit{non-unitary} CFT \cite{jianMeasurementinducedCriticalityRandom2020}, though there may be important differences between $q \to \infty$ and finite $q$. Still, that $\alpha(n)$ in this spin-$\frac{1}{2}$ hybrid-MBL system is instead well described by the formula for a unitary CFT suggests that the transition in this system is quite distinct in character from that in Haar-random circuits. Finally, we note in passing that the value of $\alpha(\infty) \approx 0.34$ in this system is far from the value of $\alpha(\infty)\approx 0.49$ observed numerically in spin-$\frac{1}{2}$ Haar-random circuits \cite{zabaloCriticalPropertiesMeasurementinduced2020,entropyNote}, but close to the value of $\alpha(n) = \frac{1}{3}$ for all $n\geq 1$ predicted in Ref.\ \cite{jianMeasurementinducedCriticalityRandom2020} for the $q \to \infty$ Haar-random circuit model with periodic boundary conditions, though as emphasized this limit is far from the system considered here. It is also somewhat close to the value of $\ln{2} \times \frac{\sqrt{3}}{\pi} \approx 0.38$ for 2+0D percolation with periodic boundary conditions \cite{cardyLinkingNumbersSelfAvoiding2000,entropyNote,jiangCriticalFirstpassagePercolation2019,skinnerMeasurementInducedPhaseTransitions2019}, but, at least in hybrid Haar-random circuits, this value is correct only for the R\'{e}nyi-0 entropy \cite{jianMeasurementinducedCriticalityRandom2020}.

\section{Discussion}

Measurement-induced entanglement transitions represent an interesting new class of phase transition which shine light on the resilience of quantum entanglement against a classicality-inducing environment. They were initially explored for systems at opposite ends of the spectrum of many-body quantum dynamics: chaotic random unitary circuits, and integrable models. In this work we have demonstrated that the nature of the measurement-induced transition in many-body localized (MBL) systems can interpolate between these two extremes, in a way which is consistent with the standard phenomenology of MBL. If the measurements are made in a basis which is scrambled by the MBL dynamics, then the transition from volume- to area-law entanglement occurs at a nonzero measurement probability $p$, similar to previously studied chaotic systems. On the other hand, if the measurements are made in a basis which remains localized by the MBL dynamics, then the volume-law collapses for any nonzero $p$. This distinction does not appear with random unitary circuits, since all local operators are scrambled in time. In MBL systems, the existence of an extensive number of local integrals of motion, the `l-bits', means that not all local operators are scrambled. Instead, only those operators which have low overlap with the l-bits are scrambled, and it is for these operators that the volume-law will persist for $0 \leq p < p_{c}$ if measurements are made in the basis of their eigenstates.

One obvious question is how the measurement-induced entanglement transition (MIT) intersects with the MBL transition. At sufficiently low disorder strength $W < W_{c}$, the Hamiltonian in \cref{eq:MBL_Hamiltonian} is chaotic \cite{palManybodyLocalizationPhase2010,doggenManybodyLocalizationDelocalization2018}, i.e.\ it satisfies the eigenstate thermalization hypothesis \cite{srednickiChaosQuantumThermalization1994,dalessioQuantumChaosEigenstate2016}. One might expect the MIT in this chaotic Hamiltonian system with short-range interactions to be in the same universality class as the MIT in random local unitary circuits with the same local Hilbert space dimension. But, from the analysis of the scaling of the critical R\'{e}nyi entropies in \cref{sec:X_basis_measurements}, it appears that the MIT in the MBL transition may be in a distinct universality class to that in random local unitary circuits. This begs the question of how the conformal field theory describing the MIT critical point changes as one sweeps the disorder strength $W$ across the MBL transition.

It is worth noting that one key difference between the model considered in this paper and random-circuit models is that here there is quenched (spatial) disorder in the unitary part of the dynamics. This is noteworthy because the critical exponent $\nu = 1.3(2)$ we extract violates a naive application of the Harris criterion $\nu \geq 2/d$ with $d=1$ \cite{harrisEffectRandomDefects1974,chayesCorrelationLengthBounds1989}. This is despite the fact that we observe $z \approx 1$ at the critical point. We speculate two possible reasons for this violation. The first is that the randomness of the measurements in both space and time means that the overall `disorder' in this problem is no longer quenched, so the Harris criterion may not apply. The second is based on the recent conjecture by Li et al.\ \cite{liConformalInvarianceQuantum2020} that the critical points of these hybrid quantum circuits are described by Euclidean CFTs, where the physical time coordinate essentially acts as a second spatial coordinate. In that sense, the Harris criterion may still apply, but with $d=2$ rather than $d=1$.

We also discussed in \cref{sec:X_basis_measurements} how X-basis measurements with small but nonzero $p$ can actually \textit{increase} the steady-state entanglement relative to $p=0$, as a consequence of the increase in diagonal entropy induced by X-basis measurements. This may be somewhat counterintuitive, given the usual picture of measurements destroying entanglement, but suggests the possibility that measurements could be used as a tool to produce states with desirable properties \cite{royMeasurementinducedSteeringQuantum2019}, such as high entanglement, in systems where simply evolving with random unitaries is not feasible.

More broadly, there is also the question of how measurements affect the characteristics of phases of pure-unitary dynamics. We have seen, such as in \cref{fig:X_basis_vN_I3_dynamics}, that at least some of the aspects of MBL phenomenology remain preserved in the volume-law phase even for nonzero $p$, such as the monotonic logarithmic growth of entanglement in time. However, it has also been argued \cite{chanUnitaryprojectiveEntanglementDynamics2019} that a steady-state volume law stable to $p>0$ necessarily implies the existence of a subleading logarithmic correction to the volume law, $S(A) \sim \alpha \ln{|A|} + s |A|$, and this has been observed in random Clifford models \cite{liMeasurementdrivenEntanglementTransition2019}, so this is one aspect which is qualitatively modified by the presence of measurements. It would be enlightening to see if such a subleading correction is also present in the steady-state of this hybrid MBL system, but this is likely out of reach of the small system sizes accessible by exact diagonalization. This may therefore be an opportunity for NISQ-era \cite{preskillQuantumComputingNISQ2018} quantum simulators to probe new physics out of the reach of numerics. Direct measurement of the entanglement entropy associated with a single quantum trajectory may be difficult, owing to the need to perform the exponentially many experimental repetitions associated with the postselection of measurement outcomes \cite{baoTheoryPhaseTransition2020} and the complexity of measuring an entropy \cite{odonnellQuantumSpectrumTesting2015,liQuantumQueryComplexity2019}. However, there have been proposals for more experimentally feasible probes of the entanglement transition based on the Fisher information \cite{baoTheoryPhaseTransition2020} and coupled ancilla qubits \cite{gullansScalableProbesMeasurementinduced2019}. It would be interesting to see if these or new techniques could be developed to allow for experimental detection of novel physics, such as the subleading logarithmic corrections to the entanglement entropy, induced by measurements of quantum systems.

\section{Acknowledgements}
We thank David Huse and Michael Gullans for enlightening discussions on their related work, and especially Michael Gullans and Adam Nahum for comments on the manuscript.
A.P.\ was funded by the European Research Council (ERC) under the European Union's Horizon 2020 research and innovation programme (grant agreement No.\ 853368). O.L.\ was supported by the Engineering and Physical Sciences Research Council [grant number EP/L015242/1].

\bibliography{bibliography}

\end{document}